\documentclass[final,5p,times,twocolumn]{elsarticle}
\usepackage{amssymb}
\usepackage{amsmath}
\usepackage{graphicx}
\usepackage{bm}
\usepackage{xcolor}
\usepackage{epsfig}
\usepackage{amsfonts}
\biboptions{sort&compress}

\newcommand{\beqa}{\begin{eqnarray}}
\newcommand{\beq}{\begin{equation}}
\newcommand{\eeqa}{\end{eqnarray}}
\newcommand{\eeq}{\end{equation}}

\begin{document}

\begin{frontmatter}

\title{Dynamics of a quantum wave emitted by a decaying and evanescent point source}
\author[cfm,dipc,iker]{F. Delgado\corref{cor1}}
\ead{fernando.delgadoa@ehu.eus}
\address[cfm]{Centro de F\'{i}sica de Materiales, Centro Mixto CSIC-UPV/EHU, Paseo Manuel de Lardizabal 5, E-20018 Donostia-San Sebasti\'an, Spain }
\address[dipc]{Donostia International Physics Center (DIPC), Paseo Manuel de Lardizabal 4, E-20018 Donostia-San Sebasti\'an, Spain }
\address[iker]{IKERBASQUE, Basque Foundation for Science, E-48013 Bilbao, Spain}

\author[qf,sha]{J. G. Muga}
\address[qf]{Departamento de Qu\'{i}mica-F\'{i}sica, UPV/EHU, Apartado 644, 48080 Bilbao, Spain }
\address[sha]{Department of Physics, Shanghai University, 200444 Shanghai, People's Republic of China}
\begin{abstract}
%
We put forward a model that describes a decaying and evanescent point source of
non-interacting quantum waves in 1D. 
This point-source assumption allows for a simple description that captures the essential aspects of the dynamics
of a wave traveling through a classically forbidden region
 without the need to specify the details of the inner region. The dynamics of the resulting wave is examined 
and several characteristic times are identified. One of them generalizes the tunneling time-scale introduced by B\"uttiker and Landauer and it characterizes the arrival of the the maximum of the wave function.
  Diffraction in time and deviations from exponential 
decay are also studied. Here we show that there exists an optimal injection frequency and detection point for the observation of these two quantum phenomena. 

\end{abstract}


\end{frontmatter}


\section{Introduction}
Since the pioneering work of B\"uttiker and Landauer (BL) \cite{Buttiker_Landauer_prl_1982,Buttiker_prb_1983}, much effort has been devoted to understand and characterize
the dynamics of wave functions representing the state of a quantum particle propagating through a tunneling region. At first the research was oriented to define a 
``tunnelling time'' for the particle. Several candidates were put forward, apart from the BL time, and an intense debate followed \cite{Hauge_Stovneng_rmp_1989,Muga_Mayato_book_2008}. It
was later understood that since the projectors for locating the particle at the barrier and for finding the particle eventually transmitted 
do not commute, many possible quantizations of the classical 
concept of a traversal time through the barrier region are possible, and that several of them may be relevant depending on the
experimental setting and/or quantity observed \cite{Brouard_Sala_pra_1994,Muga_Mayato_book_2008}. Thus, rather than seeking ``the tunneling time'', a research line aimed at 
describing the wave dynamics with a minimal number of elements emerged, which could  include characteristic times for forerunners, main peaks,
or transitions among different regimes \cite{Brouard_Muga_pra_1996,Muga_Buttiker_pra_2000,Calderon_Villavicencio_pra_2001,Gaston_Villavicencio_pra_2002,Delgado_Muga_pra_2003,Ban_Sherman_pra_2010}. 

Simplified models are instrumental in identifying the main phenomena and develop the necessary conceptual 
frame as well as a general theory.   
Among the different analytical models used to study transient phenomena in quantum mechanics \cite{Campo_Calderon_pr_2009}, the ``source models'', where the  wave function is given at a fixed position for all times, play a key role. They have been used to study diffraction in time \cite{Gahler_Golub_zpb_1984,Brown_Summhammer_1992,Felber_Muller_1990,Brukner_Zeilinger_pra_1997,Campo_Muga_jpa_2005,Torrontegui_Munoz_pra_2011}, tunneling dynamics~\cite{Stevens_jpc_1983,Ranfagni_Mugnai_ps_1990,Moretti_ps_1992,
Buttiker_Thomas_ap_1998,Buttiker_Thomas_sm_1998,Muga_Buttiker_pra_2000,Gaston_Villavicencio_pra_2002,Delgado_Muga_pra_2003},  
dynamics on absorbing media \cite{Delgado_Muga_pra_2004,Ruschhaupt_Muga_prl_2004}, 
atom lasers~\cite{Campo_Muga_jpb_2007}, deviations from exponential decay \cite{Torrontegui_Muga_pra_2009}, and the time of arrival \cite{Allcock_ap_1969,Muga_leavens_pr_2000,Baute_Egusquiza_jpa_2001}. 
The physical meaning of the source boundary conditions was clarified in \cite{Baute_Egusquiza_jpa_2001,Campo_Muga_jpa_2005} by finding 
the connection between source boundary conditions and the more standard initial value problem.  

In most  applications of the point-source model, the emission (carrier) frequency was real, for either traveling or evanescent conditions (with real or imaginary wavenumber, respectively). An imaginary part was added to the carrier frequency in \cite{Torrontegui_Muga_pra_2009}
to study deviations from exponential decay and their enhancement, and also in  \cite{Torrontegui_Munoz_pra_2011} to find a simple explanation 
of diffraction in time (DIT). In this work we consider a case so far overlooked, namely, a negative real part of the frequency  corresponding to evanescent conditions and an imaginary part that produces decay. This completes the work of one of us with M. B\"uttiker in  \cite{Muga_Buttiker_pra_2000}, 
which dealt with a purely evanescent source, without decay, and also fills the gap between this work and the decaying source considered in 
\cite{Torrontegui_Muga_pra_2009}. The physical setting corresponds to the 1D wave dynamics in an evanescent region (positions $x>0$) for a decaying 
resonance which is depleted exponentially through some escape channel (say to the left) which is not represented explicitly in the model, see Fig. \ref{scheme}(a).

A surprising result for the purely evanescent source~\cite{Muga_Buttiker_pra_2000,Villavicencio_Romo_pra_2002}, 
was that a direct generalization  of the BL time 
set a time scale for the wave density maximum 
 in opaque (semiclassical) conditions, i.e., beyond the penetration length. This ``forerunner'', 
paradoxically, was not at all dominated by evanescent components but by a saddle point 
contribution above threshold. This finding provided a role for the BL timescale different from the ones that had been attributed so far to it (as            
a scale that determines the transition from sudden to adiabatic
regimes for an oscillating barrier~\cite{Buttiker_Landauer_prl_1982}, and  the rotation of the spin in a weak magnetic field in opaque
conditions~\cite{Buttiker_prb_1983}). Here we shall generalize the BL time scale further for the decaying and evanescent source and specify its relation 
to the saddle-point dominated peak. Unlike  \cite{Muga_Buttiker_pra_2000}, the decaying evanescent source allows for a power-law decay following the commonly observed exponential one (post-exponential regime), which we analyze in this work. As well, the modifications on DIT with respect to the decaying  above-threshold source in \cite{Torrontegui_Muga_pra_2009}
are examined.   
\section{Point source model\label{intro}}
%
%
%
%
Let a source at the origin $\tilde{x}=0$, be switched on suddenly at time $\tilde t=0$.
(Dimensional quantities wear a tilde here to distinguish them from dimensionless ones, without tilde.)  
The source {\em boundary} condition is 
\beqa
\tilde \psi (\tilde x=0,\tilde t)=\Theta(\tilde t)e^{-i\tilde \omega_0 \tilde t},
\eeqa
with complex (carrier) frequency $\tilde \omega_0=\tilde \omega_0^R+i\tilde \omega_0^I$. The particle is assumed to move in one 
dimension and we consider the emission into $\tilde{x}\ge 0$.  
The initial wave function is $0$ everywhere except at  $\tilde x=0$. This setting was studied in Ref.~\cite{Torrontegui_Muga_pra_2009,Torrontegui_Munoz_pra_2011} assuming the propagating condition $\tilde \omega_0^R>0$ and $\tilde\omega_0^I<0$. 
This corresponds to a simplified model to account for the propagation into the $\tilde x\ge 0$  region of an initially prepared 
resonant state with frequency $\tilde \omega_0^ R$ above the cut-off frequency $\tilde \omega_c=0$ and lifetime $-1/\tilde\omega_0^I$. 
Notice that, without loss of generality, we fix the constant potential $V_0$ in which the quantum particle moves as zero, and hence, frequencies bellow $0$ lead to imaginary wavenumbers (evanescent waves).

By contrast, here we consider the evanescent injection below the media cut-off, i.e., $\tilde\omega_0^ R\le 0$. 
In addition, since  
$|\tilde \psi(0,\tilde t)|^2=e^{2\tilde \omega_0^I t}$, we impose that 
$\tilde\omega_0^ I\le 0$ to model an exponentially decaying source. 
%

The dispersion relation corresponding to the free particle (unlike \cite{Muga_Buttiker_pra_2000},
we set the constant potential level as $V=0$) is 
\beqa
\tilde\omega(\tilde k)=\frac{\hbar\tilde k^2}{2m}.
\eeqa
We define $\tilde{k}=\sqrt{2m\tilde\omega/\hbar}$ with a branch cut slightly below the real axis. 
For $\tilde{k}_0=\sqrt{2m\tilde\omega_0/\hbar}$  
its real part is negative or zero and the imaginary part is zero or positive.
\subsection{Dimensionless Schr\"odinger equation}
We introduce dimensionless quantities in terms of a characteristic length $L$,
\beqa
x&=&\tilde x/L,\crcr
t&=&\tilde t\frac{\hbar}{2mL^2},\crcr
\psi(x,t)&=&\sqrt{L}\tilde \psi(\tilde x,\tilde t),
\eeqa
so that the Schr\"odinger equation takes the form
\beq
i\frac{\partial \psi(x,t)}{\partial t}=-\frac{\partial^2\psi(x,t)}{\partial x^2}.
\eeq
If we define the dimensionless wavenumber $k=\tilde k L$ and frequency $\omega(k)=k^2$, we have that
\beqa
\omega_0=k_0^2=\left(k_{0,R}^2 -k_{0,I}^2\right)+2i k_{0,I} k_{0,R}
\eeqa
where $k_{0,R}={\rm Re}[k_0]$ and $k_{0,I}={\rm Im}[k_0]$. 
The conditions of below cut-off injection and decay will translate into
\beqa
k_{0,R}^2 -k_{0,I}^2 &&\le 0,
\crcr
k_{0,I} k_{0,R} && \le 0.
\eeqa
Here we choose the characteristic length $L$ as
\beqa
L=1/|\tilde k_{0,I}|.
\eeqa
In accordance with our branch-cut criterion, we take $\tilde k_{0,I}>0$ and $\tilde k_{0,R}\le 0$.  
For convenience, we introduce the (positive) dimensionless velocity $v_0=-k_{0,R}$. Thus, $k_0=-v_0+i$ and we can write
\beqa
\omega_0=\left(v_0^2-1\right)-2i v_0, \qquad 0 \le v_0 \le 1.
\label{defs}
\eeqa
%
The dimensionless velocity $v_0$ is bounded by two limits. The $v_0=0$ limit corresponds to the non-decaying source, i.e., particles are continuously injected into the system at all times with the same intensity. The $v_0=1$ limit, on the other hand, corresponds to the injection at the exact cut-off frequency of the media, i.e., a finite lifetime resonance centered at zero.
\begin{figure}
  \begin{center}
    \includegraphics[width=1.\linewidth,angle=0]{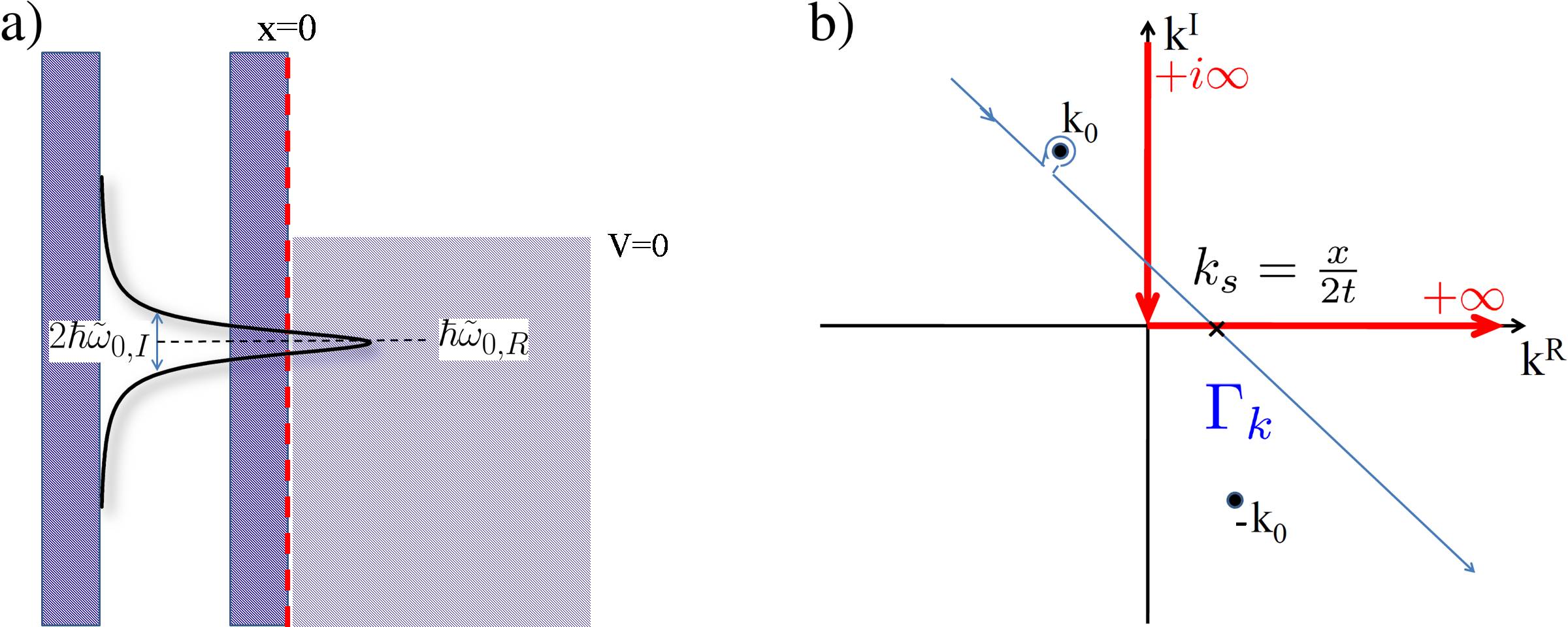} 
  \end{center}
  \caption{ Scheme of the model and details of the complex plane analysis.
{\bf (a)}  The model accounts for the wave dynamics of a decaying resonance, with central frequency $\tilde \omega_{0,R}<0$, into a tunnel region ($x>0$) that induces an exponential depletion, with a characteristic lifetime $1/(2\tilde\omega_{0,I})$, of the initial state.  
{\bf (b)}  Original integration path  $\gamma$ in the complex $k$-space (red arrows) and contour $\Gamma_k$ of integration for the $w$ function crossing the $k_s=x/2t$ saddle point and passing above all possible poles (in our case, only the pole at $k_0$ may be crossed). 
 }
\label{scheme}
\end{figure}
 \subsection{Exact solution\label{esolution}}
The bounded solution of the Schr\"odinger equation consistent with the source boundary condition may be written in the 
form \cite{Torrontegui_Muga_pra_2009} 
\beqa
\label{inte}
\psi(x,t)=\int_{-\infty}^{\infty}d\omega A(\omega)e^{i\omega^{1/2}x}e^{-i\omega t} ,
\eeqa
where
\beqa
A(\omega)=\frac{i}{2\pi(\omega-\omega_0)}.
\eeqa
The integral in Eq. (\ref{inte}) is easier to handle in the complex $k=\sqrt{\omega}$ plane.
 The branch cut for the square root is chosen as before, just below the positive real axis, which ensures a decaying solution for $x\to \infty$.
Hence, we get
\beqa
\psi(x,t)=\int_\gamma dk\frac{ik}{\pi(k^2-k_0^2)} e^{ikx-ik^2t},
\eeqa
where the path $\gamma$ goes from $k=i\infty$ to the origin and from there to $k=\infty$ along the real axis, see Fig. \ref{scheme}(b).
Using the identity
\beqa
 \frac{1}{k-k_0}+\frac{1}{k+k_0}=\frac{2k}{k^2-k_0^2},
\eeqa
we can write
\beqa
 \psi(x,t)=\frac{i}{2\pi}\int_\gamma\! dk \!\left(\frac{1}{k-k_0}+\frac{1}{k+k_0}\right)\!e^{ikx-ik^2t}.
 \label{integg}
\eeqa
This integral can be done by deforming the integration contour as shown in Fig.~\ref{scheme}(b),
\beqa
\int_\gamma dk\left(\dots\right)=\int_{\Gamma_k} dk\left(\dots\right),
\eeqa
where the contour $\Gamma_k$ goes along the steepest descent path, a straight line  $k_I=-k_R+k_s$, with the saddle point at  
$k_s=x/(2t)$, and passes above all poles. In particular $\Gamma_k$ encircles the pole at $k_0$ after the 
steepest descent path crosses it at the critical time $t_c=x/[2(1-v_0)]$.    

Now we use the following analytical result~\cite{Abramowitz_Stegun_book_1972}
\beqa
\label{wf}
{\cal I}(k_p)=\int_{\Gamma_k} dk \frac{e^{-i(tk^2-xk)}}{k-k_p}=-i\pi e^{-ix^2/(4t)}w(-u_p),
\crcr
\eeqa
where
$w(z)=e^{-z^2}{\rm Erfc}(-iz)$ is the Faddeyeva function and 
\beqa
u_p=\sqrt{\frac{t}{2}}(1+i)(k_p-x/2t).
\eeqa
Identifying $k_p$ with the poles at $k_0$ and $-k_0$ in Eq. (\ref{integg}), the wave function can be written as 
\beqa
\psi(x,t)= \frac{e^{ik_s^2t}}{2}\left[w(-u_-)+w(-u_+)\right],
\label{psiE}
\eeqa
where
\beqa
u_\pm=\pm \sqrt{\frac{t}{2}}(1+i)k_0(1\pm \tau/t),
\label{defU0}
\eeqa
and we have defined the complex time $\tau=-x/(2k_0)$. 
The modulus of $\tau$ generalizes the traversal time of B\"uttiker and Landauer \cite{Buttiker_Landauer_prl_1982,Buttiker_prb_1983}, 
and  tends to that time if decay is suppressed. Its real and imaginary parts are 
\beqa
\tau_R&=&{\rm Re}[\tau]=\frac{x v_0}{2\left(v_0^2+1\right)}  , 
\crcr\cr
\tau_I&=&{\rm Im}[\tau]=\frac{x}{2\left(v_0^2+1\right)},
\eeqa
and its modulus is
\beqa
|\tau | &=& \frac{x}{2|k_0|}=\frac{x}{2\sqrt{v_0^2+1}}. 
\eeqa
We will see in the next section that the maximum of the saddle point contribution, which is responsible for  
deviations from the pure exponential decay, arrives at $t\sim |\tau|$ (notice that $\tau_I=|\tau|$ for $v_0=0$). Furthermore, since $\tau_R=v_0\tau_I\le \tau_I$, all relevant time scales for the long-time behavior can be described in terms of a single time parameter $\tau_I\sim|\tau|$.

\section{Asymptotic behavior}
%
%
%
%
The saddle contribution can be calculated by setting $k=k_s$ in the denominators of Eq. (\ref{integg}) and performing the remaining integration (alternatively, retain the dominant non-exponential- term in the large $|z|$ expansion of $w(z)$, see, e.g. \cite{Muga_Buttiker_pra_2000}).
In so doing, one gets
\beqa
\psi_S(x,t)&=&\frac{e^{i\pi/4}}{2\sqrt{\pi}}e^{ik_S^2t}t^{-1/2}\!\left(\frac{1}{k_s-k_0}+\frac{1}{k_s+k_0}\right)\!
\nonumber
\\
&=&-\sqrt{\frac{2 t}{\pi}}\frac{e^{ik_S^2 t}}{(i-1)k_0} \frac{\tau}{t^2-\tau^2}.
\label{psis}
\eeqa
The associated probability density 
\beq
\label{psi2}
|\psi_S|^2=\frac{t|\tau|^2}{\pi |k_0|^2[t^4+|\tau|^4-2t^2{\rm{Re}}(\tau^2)]}
\eeq
has a maximum with respect to $t$, for fixed $x$, at 
\beq
t_{Max}=\frac{1}{\sqrt{3}}\bigg[\tau_R^2-\tau_I^2+2\sqrt{\tau_R^4+\tau_I^4+\tau_R^2\tau_I^2}\bigg]^{1/2}.
\eeq
Note that $t_{Max}$ depends linearly on $x$. It tends to $|\tau|/\sqrt{3}$ in the purely evanescent limit ($v_0=0$), and to $|\tau|/{3}^{1/4}$
in the opposite emission-at-threshold limit ($v_0=1$). The picture is somewhat simpler if instead of taking the derivative with respect to 
$t$ to get the maximum, we take the derivative with respect to $x$ for a given $t$ in  Eq. (\ref{psi2}), as in \cite{Villavicencio_Romo_pra_2002}. Then the maximum 
of a snapshot of the density at a given time is at 
\beq\label{xmax}
x_{Max}=2 t \sqrt{v_0^2+1}.
\eeq
In other words, the so defined peak of the density arrives at a given point $x$ at exactly $|\tau|$.

For the pole contribution at $k=k_0$, we get
 \beqa
 \psi_0(x,t)=e^{-i(v_0^2-1)t-iv_0 x} e^{-2v_0t-x}\Theta\left(t-t_c\right),
 \label{pole}
 \eeqa
 where $t_c=x/[2(1-v_{0})]$.
Contrary to the case of injection above the threshold frequency ($\omega_0>0$) studied in Refs.~\cite{Torrontegui_Muga_pra_2009,Torrontegui_Munoz_pra_2011}, the wave associated with  the pole contribution (\ref{pole}) propagates 
leftwards (with a negative wavenumber) and it decays for increasing $x$.
At a given time the positions where the pole contribute are to the left of the 
critical value $x_c=2t(1-v_{0})$. This point moves rightwards with increasing time. Incidentally, a footnote in~\cite{Torrontegui_Muga_pra_2009} 
claiming the contrary was in error.  
 
Whenever the distance between the two singular points, pole and saddle, is large, we can approximate 
\beqa
\psi(x,t)\approx \psi_0(x,t)+\psi_S(x,t).
\label{psiapp}
\eeqa
This approximation is found from the asymptotic  expansion of the $w(z)$ function  to dominant order in $1/z$ for large $|z|$~\cite{Abramowitz_Stegun_book_1972}. The required condition for such an expansion to hold is
\beqa
|u_{\pm}|=|k_0|\sqrt{x}\left(\frac{t}{2|\tau|}\pm \frac{\tau_R}{|\tau|} +\frac{|\tau|}{2t}\right)^{1/2}\gg 1,
\eeqa
which may occur for values of $t$ much smaller or larger than $|\tau|$. The actual maximum of the probability density $|\psi(x,t)|^2$ occurs at times that can differ slightly from the the maximum of the saddle contribution $t_{{\rm Max}}$, as observed for instance in Fig. \ref{figJ}b), revealing again that the approximation provided by Eq. (\ref{psiapp}) is not accurate for $t\sim\tau$.  
%
\subsection{Flux normalization\label{Fnorma}}

In order to compare the results corresponding to different injection frequencies, 
we normalize the wave functions to have the same number of emitted particles (one) in all cases. 
This is done introducing the normalized function
\beqa
\psi_N(x,t)=\left[ \int_0^ \infty dt J(x=0,t)\right]^{-1/2}\times\psi(x,t),
\label{psiNorm}
\eeqa
where the flux of particles $J(x,t)$ is defined as
\beqa
J(x,t)=2 {\rm Im}\left[\psi^*(x,t)\frac{\partial \psi(x,t)}{\partial x}\right].
\eeqa
The flux at the origin decays on a typical time scale $t\sim 2v_0$ but it can oscillate around zero reaching negative values for some time intervals, see Fig.~\ref{figJ}(a). However, the time integral in Eq. (\ref{psiNorm}) is always positive. The inset of Fig.~\ref{figJ}(a) shows the momentum dependence of the normalization factor in Eq. (\ref{psiNorm}). 
%
%
%
%
%
%
\section{Results\label{results}}
\begin{figure}
  \begin{center}
    \includegraphics[height=0.8\linewidth,angle=0]{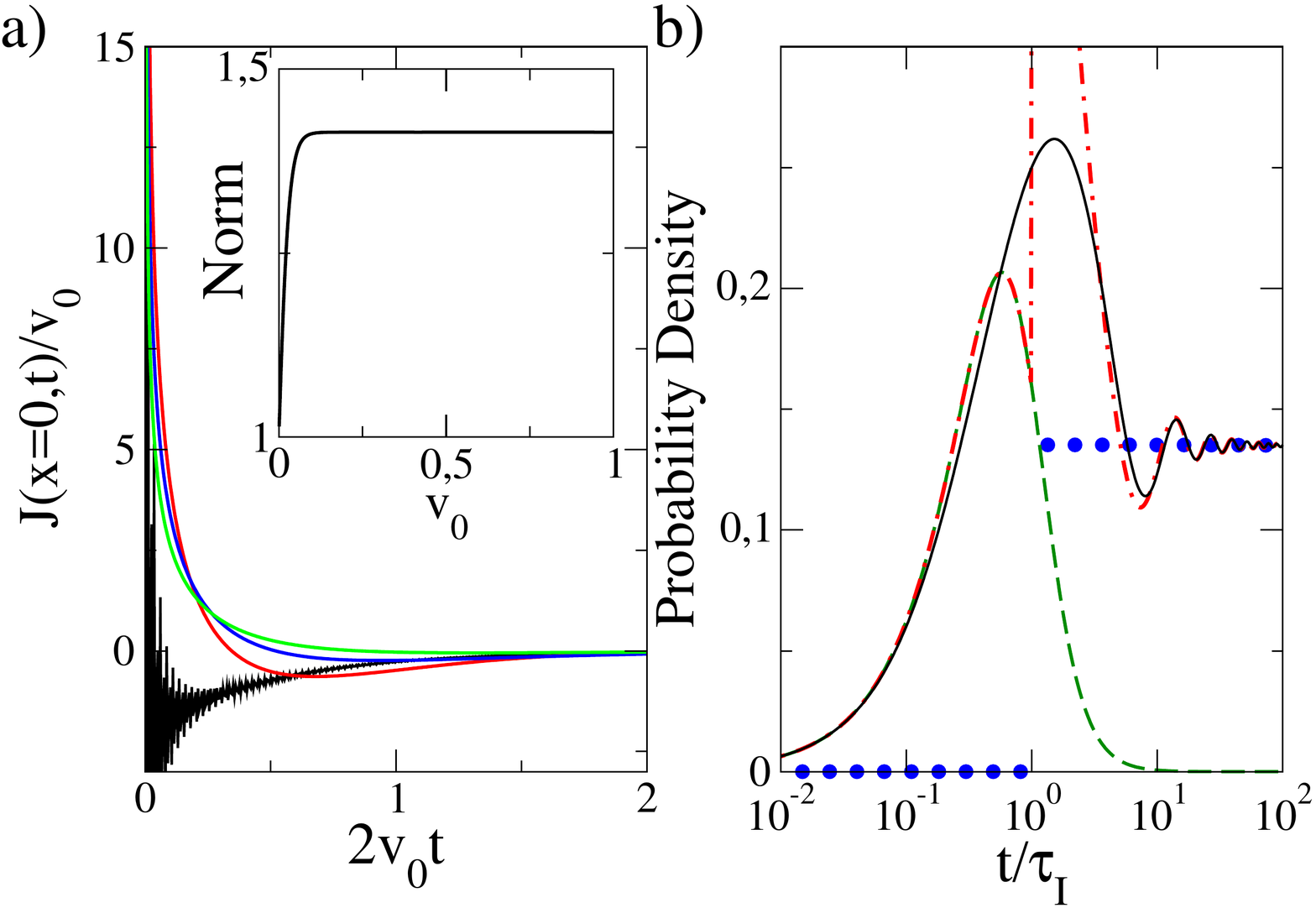} 
  \end{center}
  \caption{Temporal dependence of the wave function.
  {\bf (a)} Temporal dependence of the flux at $x=0$ for four different values of $v_0$: $10^{-3}$ (black), $0.25$ (red), $0.5$ (blue) and $0.999$ (green). Notice that, as the flux $J$ scales with the velocity, the ratio $J/v_0$ is nearly independent of $v_0$ for $v_0\sim 1$. The inset shows the normalization factor introduced in Eq. (\ref{psiNorm}).   {\bf b)} Temporal dependence of the normalized probability density at $x=1$ for $v_0=0$ (black-solid line), saddle contribution (green-dashed line), pole (blue circles) and approximate solution, Eq. (\ref{psiapp}), (red dotted-dashed line). 
  }
\label{figJ}
\end{figure}

A typical temporal dependence of the probability density $|\psi(x,t)|^2$ at short distances ($x=0.1$) for the pure evanescent wave injection ($v_0=0$) is shown in 
Fig.~\ref{figJ}(b), together with the saddle and pole contributions. 
As observed, the short time behavior $t \ll |\tau|$ (if $v_0=0$,  $|\tau|=\tau_I$) 
is reproduced solely by the non-exponential saddle contribution $|\psi_S|^2$. We may view this regime as a {short-time deviation from exponential decay}~\cite{Khalfin_sjetp_1958,Khalfin_dak_1957}.

The maximum of the saddle contribution $|\psi_S|^2$ occurs at $t_{\rm Max}\sim |\tau|$. 
 At the critical time $t_c$, the pole contribution enters, although the approximation (\ref{psiapp}) may not provide a satisfactory description for $t\sim t_c$ 
close to the source.
In the limiting case ${\rm Im}(\omega_0)=0$,  the pole contribution $|\psi_0|^2$ remains constant at $t>t_c$. Thus,  as $|\psi_S|^2$ decays, no extra features appear in the dynamics of the probability density.

For $v_0>0$, however, the pole contribution decays exponentially with a characteristic lifetime $\tau_0=1/(2v_0)$ whereas the saddle contribution decays as a power law, so there may be a transition to a post-exponential regime dominated by the saddle contribution. This is analyzed in Sec.~\ref{pexp}.  
Furthermore, when pole and saddle contributions coexist and are of similar magnitude, interferences between these two terms occur. This will be 
discussed in Sec. \ref{DIT}.
\subsection{Long times deviations from  exponential decay\label{pexp}}
\begin{figure}[t]
  \begin{center}
    \includegraphics[height=0.8\linewidth,angle=0]{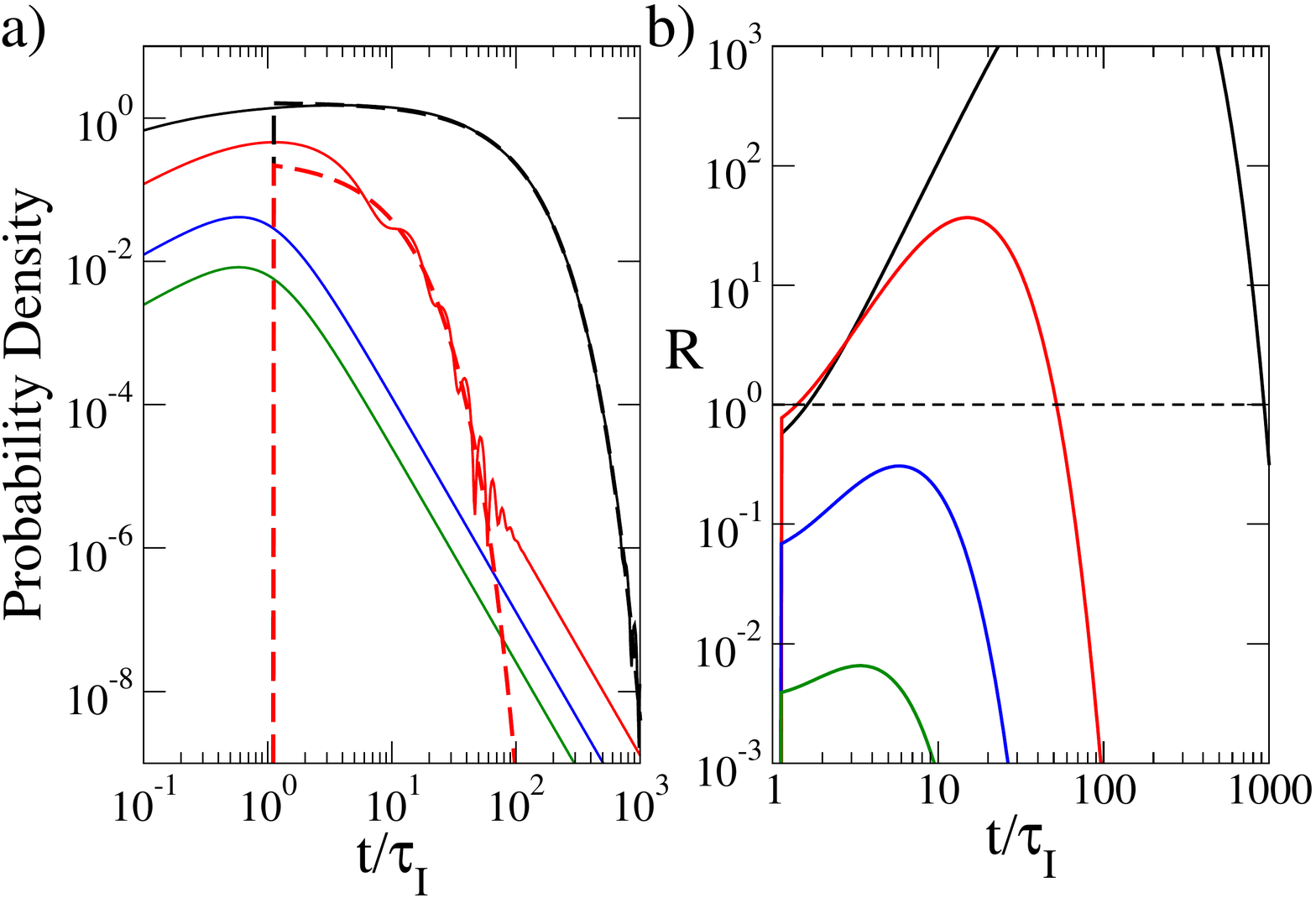} 
  \end{center}
  \caption{ Comparison between saddle and pole contributions.
  {\bf (a)} Time evolution of the normalized probability  density for $v_0=0.1$ at $x=0.1$ (black-solid line), $x=1$ (red -solid line), $x=10$ (blue), and $x=50$ (green). The dashed lines correspond to the pole contribution at $x=0.1$ (black) and $x=1$ (red), while for $x\geqslant  10$, the pole contribution is out of scale. 
  {\bf (b)} Ratio $R=|\psi_0/\psi_S|^2$ vs. time for $v_0=0.1$ at $x=0.1$ (black), $x=1$ (red), $x=2.5$ (blue) and $x=4$ (green). The horizontal line at $R=1$ is shown as a reference.   
  }
\label{fig1}
\end{figure}

%
%
%
%
%
 { Figure \ref{fig1}(a) shows a typical evolution in a regime of parameters where 
the transition to a post-exponential regime is apparent. 
Contrary to the case depicted in Fig.~\ref{figJ}(b), now the 
exponential decay given by the pole contribution 
is followed by a post-exponential one, where  the saddle contribution dominates again giving place to the {\em long-time deviations from the exponential decay}~\cite{Fonda_Ghirardi_rpp_1978,Nikolaev_pu_1969}. }
Notice that the transition can be observed only at short distances from the source ($x\lesssim 1$) since at larger distances the pole contribution is always much smaller than the saddle one.

We center our attention in the longest time crossing between the pole and saddle contributions. If we define the ratio $R=|\psi_0/\psi_S|^ 2$, the transition time $t_p$ to the post-exponential regime will be the longest
 time where $R=1$, such that the saddle contribution, leading to a power law decay, dominates for  $t>t_p$. The characteristic time $t_p$ is thus defined by the transcendental equation $R(x,t_p;v_0)=1$.
Figure \ref{fig1}(b) shows the temporal dependence of $R$ for different values of $v_0$. Due to the different scaling of the pole and saddle contributions, the product $v_0x$ does not uniquely determine the transition, and both, $v_0$ and $x$  are needed to specify $t_p$.

{According to the number of crossings with the $R=1$ line, one can find three different scenarios: i) no crossing (dominance of the saddle contribution for all times), ii) a single crossing corresponding to long times deviations, and iii) two crossings, one for short times and the other one for long times.  When the injection frequency is close to the the cut-off ($v_0\approx 1$), the pole contribution enters at $t_c\sim x/2(1-v_0)\to \infty$.  Thus, in this limit there is in general no crossing with the $R=1$ line except when $x\to 0$. By contrast, close to the non-decaying injection ($v_0=0$), one can easily find the two aforementioned scenarios as observed in Fig.~\ref{fig1}(b). For $v_0x \sim 1$ and $v_0 \lesssim 0.4$, there is not any crossing (blue and green lines), while in the opposite case, there are two crossings (red and black lines).
}


%
%
%
\begin{figure}
  \begin{center}
    \includegraphics[height=0.8\linewidth,angle=0]{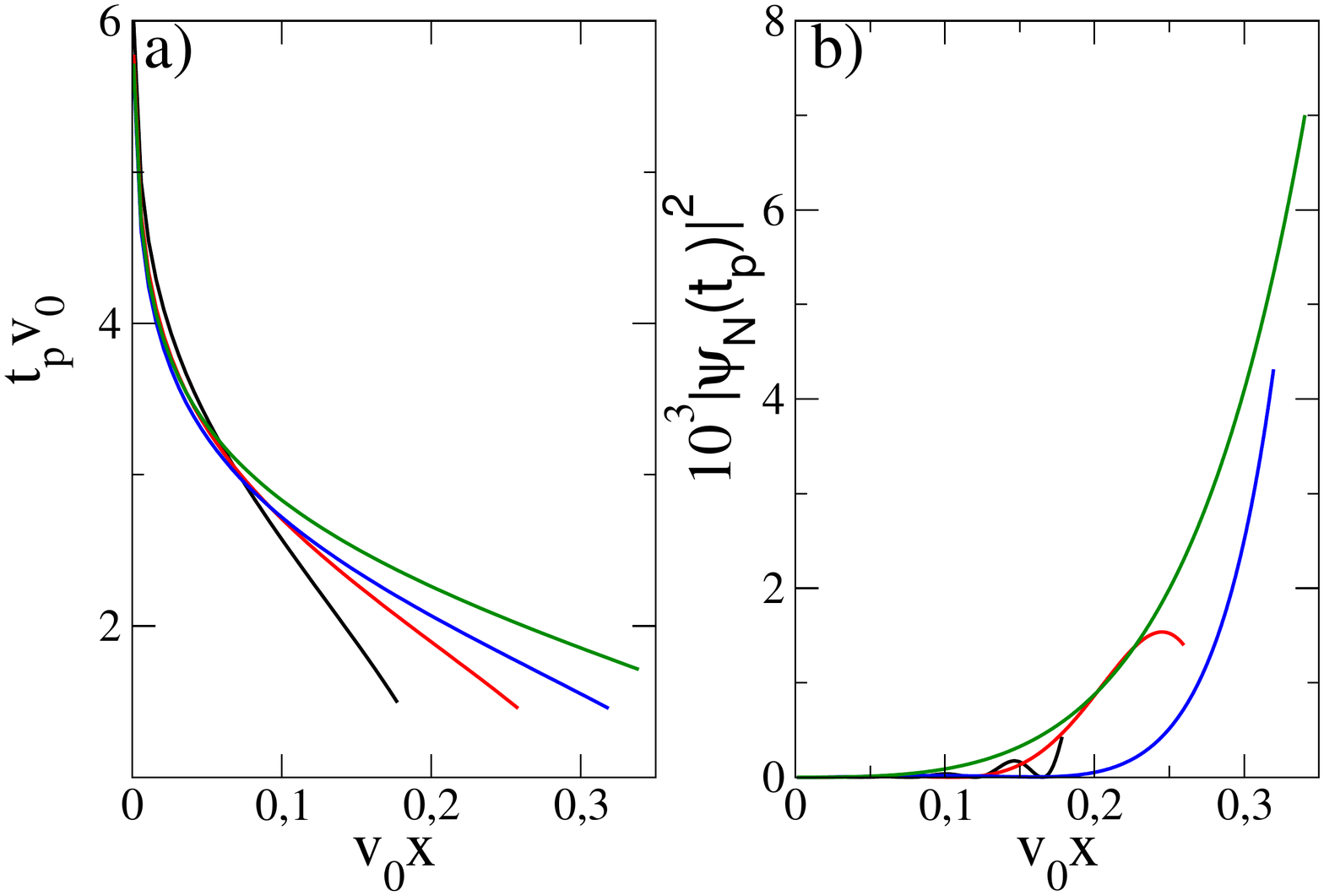} 
  \end{center}
  \caption{ Transition to the post-exponential regime.
  {\bf (a)} Spatial variation of the characteristic time $t_p$ defining the transition to the post-exponential decay with position for $v_0=0.1$ (black), $0.25$ (red), $0.5$ (blue), and
   $0.9$ (green). {\bf (b)}    Spatial dependence of the normalized probability density at $t=t_p$ for the same values of $v_0$.
  }
\label{fig2}
\end{figure}

Figure \ref{fig2}(a) shows the dependence of $t_p$ with the position. $t_p$ can be only defined for small values of $v_0x$ and it approximately scales as $1/v_0$. This would correspond to the  propagation of a classical free particle with velocity proportional to $v_0$. In addition, the further we are from the source, the more it deviates from this simple law.

An even more relevant quantity from the experimental point of view is the decay of the probability density at the transition time. As mentioned in the introduction, the main problem for the observation of long time deviations is the strong suppression of the signal. Figure \ref{fig2}(b) shows the spatial dependence of the probability density at the transition time $t_p$. Interestingly, the further from the source, the larger  the probability density is,  but bounded by the entrance of the pole contribution, i.e., $t_p>\frac{x}{2(1-v_0)}$.

%
%
%
%
%
\subsection{Diffraction in time (DIT)\label{DIT}}
\begin{figure}
  \begin{center}
    \includegraphics[height=0.8\linewidth,angle=0]{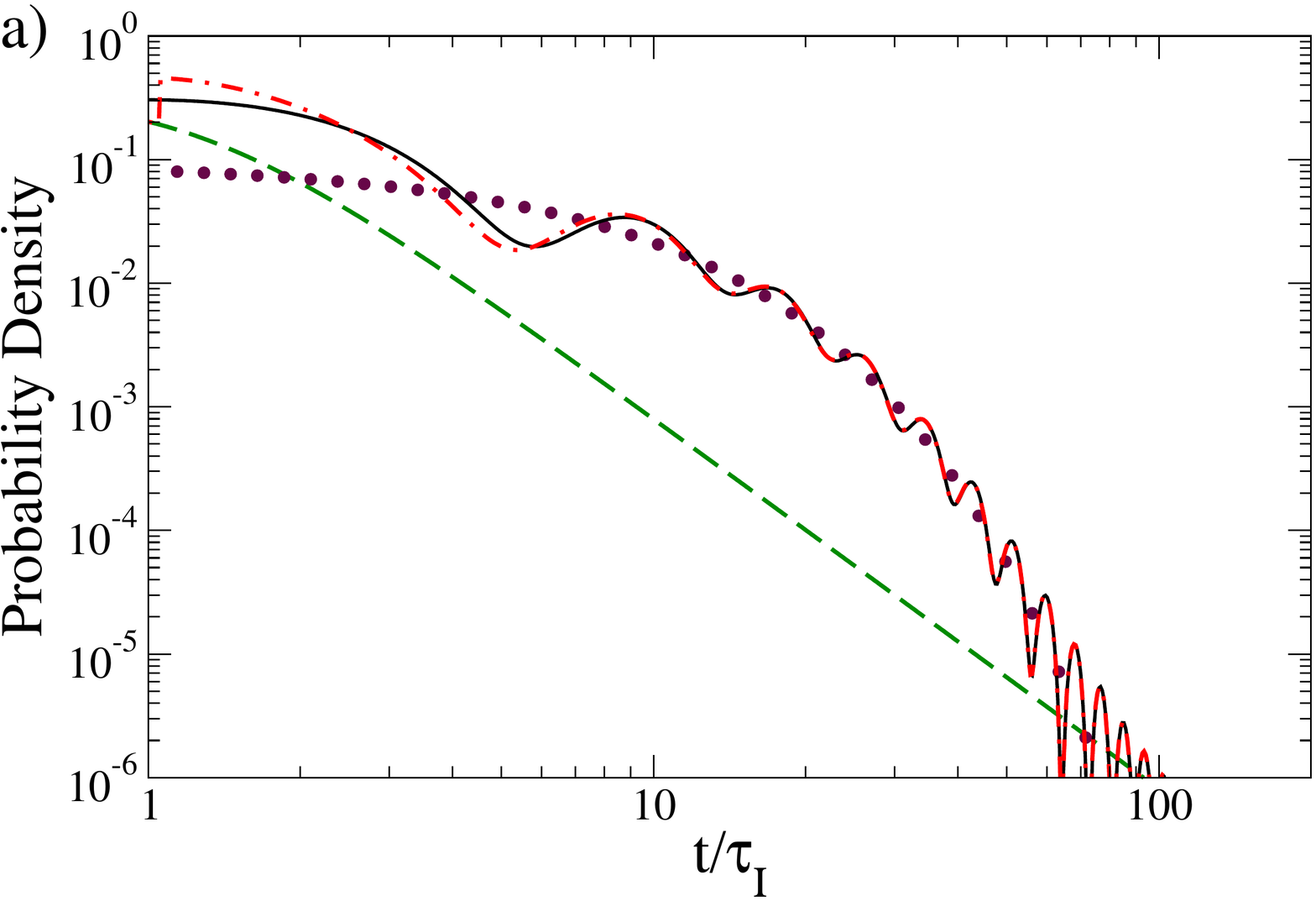} 
      \includegraphics[width=1.\linewidth,angle=0]{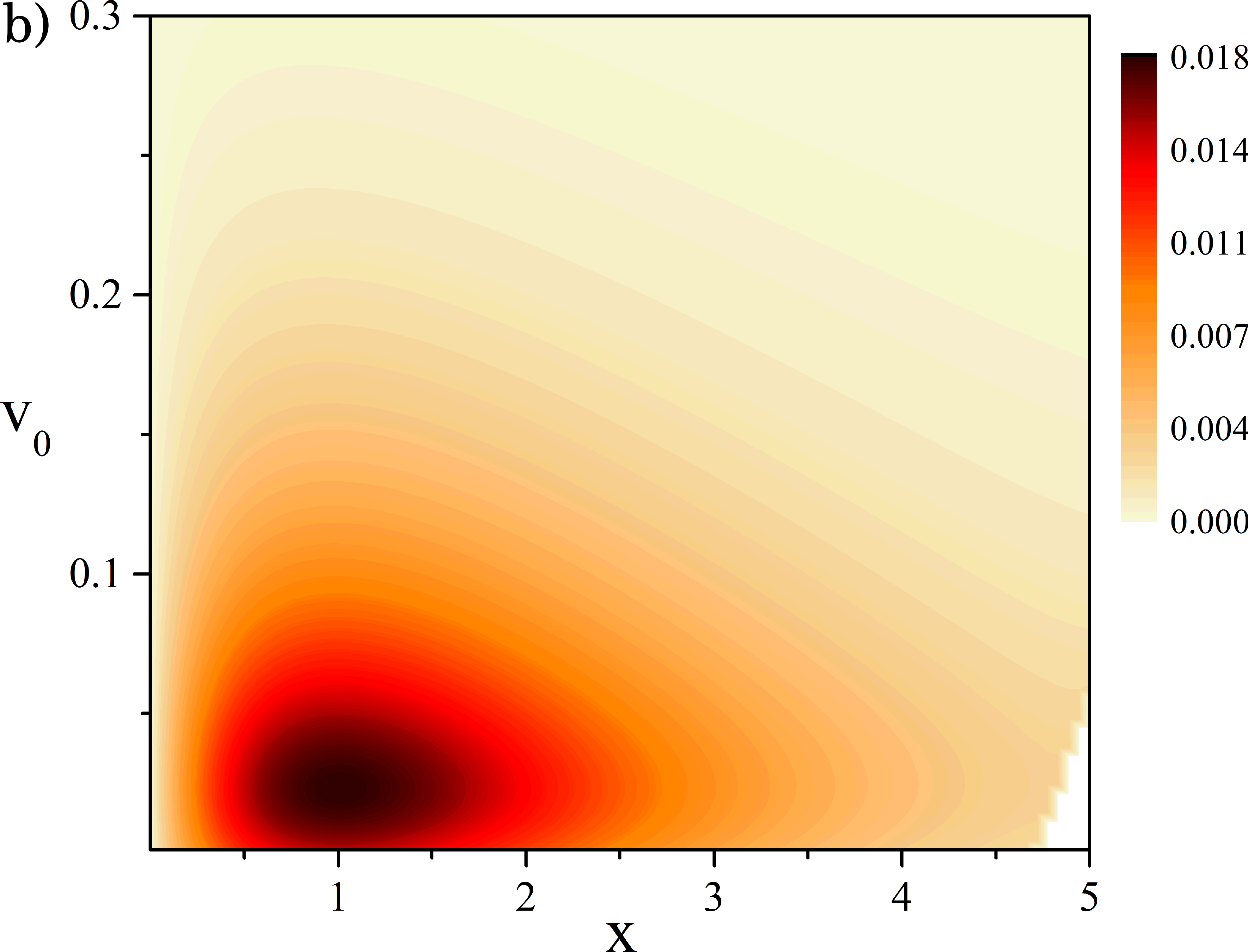} 
  \end{center}
  \caption{ Diffraction in time.
  {\bf a)} Time evolution of the exact probability density $|\psi_N|^2$ (black-solid line), saddle contribution (green dashed line), pole contribution (maroon dots), and approximate solution based on Eq. (\ref{psiapp}), (red dotted-dashed line). In all cases, $x=1.5$ and $v_{0}=0.05$.
{\bf (b)} Normalized amplitude of $\psi_{\rm Int}$ at the first minimum ($\Omega t=3\pi/2$), see Eq. (\ref{rhoint}).
  }
\label{fig4}
\end{figure}

In this section we tackle the appearance of temporal interference patterns~\cite{Moshinsky_pr_1952}. This peculiar behavior appears only in a particular regime of parameters, as seen for instance in Fig. \ref{fig1}(a). The origin for  diffraction in time (DIT) can be traced back to the interference 
between saddle and pole contributions~\cite{Torrontegui_Munoz_pra_2011}. If the exact wave function can be approximated by Eq. (\ref{psiapp}) we have that
\beqa
\left|\psi\right|^ 2\approx |\psi_S|^ 2+|\psi_0|^ 2+2{\rm Re}\left[\psi_0\psi_S\right].
\label{rhoapp}
\eeqa
As already shown in Fig. \ref{fig1}(a), the first two terms in Eq. (\ref{rhoapp}) do not show oscillations in time, so DIT is associated with the cross term $\psi_{{\rm Int}}(x,t)= 2{\rm Re}\left[\psi_0\psi_S\right]$. The visibility of the interference pattern is maximum when the amplitude of the two waves is equal. As discussed in Sec. \ref{pexp}, the condition of $|\psi_0/\psi_S|^ 2=1$ can in general occur at two different times and hence, we can find oscillations for both, at {\em short times}, soon after the maximum of the saddle contribution, and at {\em long times}, for $t\sim t_p$, although, of course, the 
short time is more relevant as the signal is stronger.  The appearance of DIT oscillations is illustrated in Fig. ~\ref{fig4}(a).

Let us now analyze the interference term $\psi_{\rm Int}$, which can be written as
\beqa
\psi_{{\rm Int}}(x,t)&=&  \frac{x}{t^{3/2} \Delta}\sqrt{\frac{1}{2\pi }} e^{-\Gamma t} (F_+ \cos (\Omega t )+F_-
   \sin (\Omega t ))
   \crcr
   &&\qquad \times
   \Theta\left(1-v_0-k_S\right)
\label{rhoint}   
\eeqa   
where $\Gamma=2(v_0+k_S)$, 
\beqa
\label{Omega}
\Omega &=& 1+k_S^2-v_{0}^2-2 k_S v_{0},
\\
\label{FRI}
F_{\pm} &=& \pm 1-v_{0}(2\pm v_{0})\pm k_S^2,
\eeqa
and
\beqa
\Delta= v_{0}^4-2 v_{0}^2
   \left(k_{S}^2-1\right)+\left(k_{S}^2+1\right)^2.
  \label{Delta}
\eeqa
Due to the exponential decay $e^{-\Gamma t}$, it is clear that $v_{0}\to 0$ enhances the visibility of the DIT. Thus, DIT is easier to observe close to the first crossing with the $R=1$ line than to the second crossing at $t_p$. 
 At the same time, due to the term $k_St=x/2$ in $\Gamma t$, the interference term also decays exponentially with $x$ even for $v_{0}=0$. This is in clear contrast to the case of a propagating wave ($\omega_0>0$), where  DIT is better observed far from the source since the signal is not exponentially suppressed with distance~\cite{Torrontegui_Munoz_pra_2011}.

Although Eq. (\ref{rhoint})  demonstrates the presence of an oscillatory term and an exponential decay, both frequency and decay rates are rather complex, depending on the position $x$ and time $t$ thought $k_S$. To get some further insight into the oscillation frequency and amplitude, since DIT occurs after the maximum of the saddle contribution at $t_{\rm Max}$, we can use the asymptotic expression for $t\gg |\tau|$. In addition, as the visibility of the DIT oscillations is enhanced for small values of $v_{0}$, we can make use of a series expansion around $v_{0}=0$. Therefore, to order $v_{0}^2$ and $(\tau_I/t)^2$,
\beq
 -\Gamma t \approx -x\left(1+v_{0}\theta\right),
 \label{expon}
\eeq
and
 \beq
 \Omega \approx 1-v_0\left(v_0+\frac{v_0}{\theta}\right),
 \label{omapp}
 \eeq
where we have defined $\theta=t/\tau_I\approx 2t(1+v_{0}^2/2)/x\gg 1$. Thus, neglecting 
higher order terms in $1/\theta$ and $v_0$, the amplitude of the oscillation is
\beq
{\cal A}\approx \frac{e^{-\Gamma t}}{\sqrt{2\pi x}\theta^{3/2}}
\left[1+2v_{0}\left(1-\frac{9}{8}v_{0}\right)\right] .
\label{ampl}
\eeq 
From Eqs. (\ref{omapp}) and (\ref{ampl}) we see that the visibility of the DIT
requires two conditions:
  first, the amplitude ${\cal A}$ should be large enough, which will occur only if the exponent in Eq. (\ref{ampl}) is small, i.e.,  $x(1+2v_0\theta)\lesssim 1$; second,
the oscillation period $T=2\pi/\Omega$ should be short enough compared to the decay time $1/\Gamma$. 

The interplay between these two different tendencies produces a non-trivial behavior of the DIT amplitude. This is illustrated in Fig. \ref{fig4}(b), where  we show the momentum and spatial dependence of the amplitude of $\psi_{\rm Int}$ at the first temporal minima of the DIT, which occurs when $\Omega t=3\pi/2$ [the probability density is normalized as given by Eq. (\ref{psiNorm})]. Importantly, DIT can not be observed for very short distances $x$ from the source, and there exists a nearly $v_0$-independent optimal detection point. 
Interestingly, although the decaying factor $e^{-\Gamma t}$ appearing in the interference term $\psi_{Int}(x,t)$ is minimized for $v_0=0$, the amplitude of the first DIT oscillation is maximized at a non-zero $v_0$.

\section{Discussion and Conclusions\label{fin}}
In this work we have analyzed the model of an evanescent and decaying point source, completing the gap between the two opposite limits studied 
before, the purely evanescent (non decaying) source~\cite{Muga_Buttiker_pra_2000},  and the source of a propagating and decaying wave~\cite{Torrontegui_Muga_pra_2009}. This model simulates the dynamics of a one-dimensional evanescent wave that originates from the decay of a finite lifetime resonance. The advantage is then that it does not depend on the specifics of the resonance region or the associated depletion channels.

A generalized B\"uttiker-Landauer time scale plays a major role as it determines the arrival of the maximum of the forerunner.  
We have demonstrated that the evanescent and decaying scenario also displays two of the main (and experimentally elusive) features of quantum decaying systems: deviations from the pure exponential decay at both short an long times and diffraction in time. Specifically, we notice that 
contrary to the ubiquitous short-time deviations, the 
 appearance of a post-exponential regime requires a finite decay time of the source, given by $1/\left|2\tilde \omega_0^I \right|$, with $\tilde \omega_0^I<0$ the imaginary part of the injection frequency. Moreover, the amplitude of the wave function at the characteristic time $\tilde t_p$ defining the transition from the exponential decay to a long-time power-law decay is maximized for injection at the exact cut-off frequency, with a maximum visibility reached for $\tilde x\approx 0.3 {\rm Im}\left[\sqrt{2m |\tilde \omega_{0,I}|/\hbar}\right]$. The small amplitude of the probability density at the transition time $\tilde t_p$, which reaches $10^{-2}m(\tilde |\tilde \omega_0|-\tilde\omega_0^ R)/\hbar$ at most,  makes the experimental detection of this regime quite challenging.

Diffraction in time~\cite{Moshinsky_pr_1952} can also appear in the case of the decaying and evanescent point source. Notably, by looking at the amplitude of the probability density at the first minimum in the DIT patterns, we have discovered that there exists an optimal region of detection points for $\tilde x\sim 0.14\sqrt{2m |\tilde\omega_0|/\hbar} $
and injection frequencies $\tilde \omega_0\approx|\tilde\omega_0|\left(-1-0.02i\right)$ that maximizes the amplitude of the DIT oscillations.

Finally we point out that in many experimental setups, in particular in solid state devices, current instead of probability density is detected. Therefore, instead of looking at the probability density, a measurement based on current detection will probe the flux at a given position and averaged over a detection time. Hence, both long time deviations and DIT may be also observed in the current density.
\section*{Acknowledgements}
{This work was supported by 
the Basque Country Government (Grant No.
IT472-10),
Ministerio de Econom\'{i}a y Competitividad (Grant No.
FIS2012-36673-C03-01),  and the program UFI 11/55 of UPV/EHU}


\end{document}